\documentclass[aps,prl,twocolumn,showpacs,
tightenlines,superscriptaddress,
preprintnumbers]{revtex4}

\def\lsim{\raise0.3ex\hbox{$<$\kern-0.75em\raise-1.1ex\hbox{$\sim$}}}
\def\gsim{\raise0.3ex\hbox{$>$\kern-0.75em\raise-1.1ex\hbox{$\sim$}}}

\usepackage{graphicx}

\begin{document}

\title{Perturbative QCD Calculations of Elliptic Flow and Shear Viscosity\\
in Au+Au Collisions at $\sqrt{s_{NN}}=200$ GeV}

%\author{Zhe Xu$^1$, Carsten Greiner$^1$,
%and Horst St\"ocker$^{1,2}$}
%\affiliation{$^1$Institut f\"ur Theoretische Physik, 
%Goethe-Universit\"at Frankfurt, Max-von-Laue-Strasse 1,
%D-60438 Frankfurt am Main, Germany\\
%$^2$Gesellschaft f\"ur Schwerionenforschung mbH (GSI),
%Planckstrasse 1, D-64291 Darmstadt, Germany}

\author{Zhe Xu}
\author{Carsten Greiner}
\affiliation{Institut f\"ur Theoretische Physik, 
Goethe-Universit\"at Frankfurt, Max-von-Laue-Strasse 1,
D-60438 Frankfurt am Main, Germany}
\author{Horst St\"ocker}
\affiliation{Gesellschaft f\"ur Schwerionenforschung mbH (GSI),
Planckstrasse 1, D-64291 Darmstadt, Germany}

\date{\today}

\begin{abstract}
The elliptic flow $v_2$ and the ratio of the shear viscosity over
the entropy density, $\eta/s$, of gluon matter are calculated from
the perturbative QCD (pQCD) based parton cascade Boltzmann approach
of multiparton scatterings. For Au+Au collisions at
$\sqrt{s}=200$A GeV the gluon plasma generates large $v_2$ values
measured at the BNL Relativistic Heavy Ion Collider. Standard pQCD 
yields $\eta/s\approx 0.08-0.15$ as small as the lower bound found from
the anti-de Sitter/conformal field theory conjecture.
\end{abstract}


\maketitle

The large values of the elliptic flow $v_2$ measured at
the Relativistic Heavy Ion Collider (RHIC) \cite{rhicv2}
indicate that the matter created, the quark gluon plasma (QGP), behaves
as a nearly perfect fluid exhibiting
strong ``explosive'' collective motion. As transport coefficients like
the shear viscosity characterize the dynamics of the QGP \cite{T03},
it is obvious that a less viscous medium will generate larger $v_2$
values. Now, ideal hydrodynamical calculations have reproduced the
observed $v_2$ dependence on centrality (within $30\%$), mass, and
transverse momentum $p_T$ (for $p_T < 1.5$ GeV) \cite{HKHRV01}. Hence,
the QGP created possesses a small viscosity coefficient \cite{L07}.
The smallness of the viscosity is of great interest as a result of the 
recent debate about speculative ``realizations'' of supersymmetric
representations of Yang-Mills theories using the anti-de Sitter/conformal
field theory (AdS/CFT) conjecture \cite{adscft}. On the other hand, 
calculations done within ideal hydrodynamics failed to reproduce 
the $v_2$ at higher centrality, at higher transverse momentum, and its 
dependence on rapidity \cite{HT02}. Although recent calculation within
viscous hydrodynamics \cite{RR07} can fit the centrality dependence of
$v_2$, the $p_T$ dependence cannot be obtained consistently. One drawback
in these calculations is the assumed ideal thermal initial conditions.
In principle, the entire process of the nuclear collision is theoretically
better studied by using a consistent microscopic transport theory.
Transport simulations can provide the correct viscous hydrodynamics
when the system is near equilibrium. Moreover, nonequilibrium processes,
like equilibration and the onset and the breakdown of viscous hydrodynamics,
are involved. However, both hadronic \cite{BS02} and partonic \cite{MG02}
cascade calculations with $2\to 2$ processes failed to yield the large
$v_2$ values observed at RHIC.

Recently we have developed a relativistic perturbative QCD (pQCD) based
on-shell parton cascade Boltzmann approach of multiparton scatterings
(BAMPS) \cite{XG05,XG07} to study, on a semiclassical
level, the dynamics of gluon matter produced in Au+Au collisions at RHIC
energies. Although coherent quantum effects like color
instabilities \cite{instab} may play a role at the very initial stage where
the matter is superdense, it is the pQCD interaction that drives the
gluon matter to equilibrium within a time interval of $1$ fm/c. The fast
thermalization observed is due to gluon bremsstrahlung \cite{XG07}.
This process is also responsible for the small ratio of the shear viscosity
to the entropy density, $\eta/s$, for a static gluon gas \cite{XGviscos}.
For a coupling constant $\alpha_s=0.6(0.3)$ we obtain $\eta/s=0.076(0.13)$,
which is as small as the lower bound $\eta/s=1/4\pi$ found from the
AdS/CFT conjecture  \cite{adscft}. Hence, pQCD interactions can explain
why the QGP created at RHIC is strongly coupled and behaves like a nearly
perfect fluid. There is no need to invoke exotic black hole physics in
higher dimensions and supersymmetric Yang-Mills theories using the AdS/CFT
conjecture to understand RHIC data.

In this Letter we employ BAMPS to calculate the elliptic flow $v_2$ of
gluons in Au+Au collisions at $\sqrt{s}=200$A GeV. For a firm footing
we compare our results with the experimental data, assuming parton-hadron
duality. To see how viscous the gluon plasma behaves, the ratio of shear
viscosity to entropy density is calculated.

The initial gluon distributions are taken as given in Ref. \cite{XG07}:
an ensemble of minijets with transverse momenta greater than $1.4$ GeV, 
produced via semihard nucleon-nucleon collisions. We use Glauber geometry
with a Woods-Saxon profile and assume independent binary nucleon-nucleon
collisions. A formation time for initial minijets is also
included \cite{XG05,XG07}. In contrast to the Glauber-type initial
condition, the color-glass-condensate--type initial condition provides
larger initial eccentricity and thus might yield larger $v_2$ values
\cite{dumi}. Including quarks in BAMPS is in progress. We do not expect large
changes in the present results shown below, because the quark amount at
RHIC is initially only $20\%$. As quarks are less interactive than gluons,
the total $v_2$ will be slightly smaller than that obtained in a pure
gluon matter.

Gluon interactions included in BAMPS are elastic scatterings with
the pQCD cross section
$d\sigma^{gg\to gg}/dq_{\perp}^2=9\pi\alpha_s^2/(q_{\perp}^2+m_D^2)^2$,
as well as pQCD inspired bremsstrahlung
$gg\leftrightarrow ggg$ with the effective matrix element \cite{biro}
\begin{eqnarray}
\label{m23}
| {\cal M}_{gg \to ggg} |^2 &=&\frac{9 g^4}{2}
\frac{s^2}{({\bf q}_{\perp}^2+m_D^2)^2}\,
 \frac{12 g^2 {\bf q}_{\perp}^2}
{{\bf k}_{\perp}^2 [({\bf k}_{\perp}-{\bf q}_{\perp})^2+m_D^2]}
\nonumber \\
&&\Theta(k_{\perp}\Lambda_g-\cosh y)\,,
\end{eqnarray}
where $g^2=4\pi\alpha_s$. ${\bf q}_{\perp}$ and
${\bf k}_{\perp}$ denote the perpendicular component of the momentum
transfer and of the radiated gluon momentum in the center-of-mass
frame of the collision, respectively. $y$ is the momentum rapidity of
the radiated gluon in the center-of-mass frame, and $\Lambda_g$ is the
gluon mean free path, which is calculated self-consistently \cite{XG05}.
The interactions of the massless gluons are screened by a Debye mass
$m_D^2=\pi d_G \,\alpha_s N_c \int d^3p /(2\pi)^3 \cdot f / p$,
where $d_G=16$ is the gluon degeneracy factor for $N_c=3$.
$m_D$ is calculated locally using the gluon density function $f$
obtained from the BAMPS simulation. The suppression of the bremsstrahlung
due to the Landau-Pomeranchuk-Migdal effect is taken into account
within the Bethe-Heitler regime using the step function
in Eq. (\ref{m23}). 

In the present pQCD simulations the interactions of the gluons are
stopped when the local energy density
drops below $1\ \rm{GeV/fm}^3$. This value is assumed to be the critical
value for the occurrence of hadronization, below which parton dynamics is
not valid. Because hadronization and hadronic cascade are
not yet included in BAMPS, a gluon, which ceases to interact, propagates
freely and can be regarded as a free pion according to the
parton-hadron duality.

In Table \ref{table1} the initial and final transverse energy per rapidity
$dE_T/dy(|y| < 1.5)$ for $6$ centrality bins are listed, which are
obtained from BAMPS using two different coupling constants.
\begin{table}[b]
\begin{center}
\caption{
\label{table1}
Centrality dependence of the transverse energy per rapidity at midrapidity.
Dimension is GeV.}
\begin{tabular}{c c c c c}
\hline 
\hline 
Centrality & PHOBOS & Initial & Final $\alpha_s=0.3$
& Final $\alpha_s=0.6$ \\
\hline
45\%-50\% &  58  &   85 &  61  &  55\\
35\%-45\% &  85  &  132 &  89  &  80\\
25\%-35\% & 132  &  220 & 138  & 122\\
15\%-25\% & 198  &  354 & 209  & 182\\
6\%-15\%  & 281  &  535 & 299  & 257\\
0\%-6\%   & 368  &  707 & 380  & 324\\
\hline
\hline
\end{tabular}
\end{center}  
\end{table}

The PHOBOS values are extracted from the $p_T$ spectra
for charged hadrons \cite{phobos1}, i.e.,
$dE_T=\int dp_T\, \sqrt{p_T^2+m_0^2}\times$ $dN/dp_T$ with $m_0$ being the pion 
mass. To compare with them we have multiplied a factor of $2/3$ to 
$dE_T/dy$ for gluons. The $dE_T/dy$ using $\alpha_s=0.3(0.6)$ is
slightly larger(smaller) than the PHOBOS data within $\pm 8\%$.

\begin{figure}[b]
\begin{center}
\includegraphics[height=5.4cm]{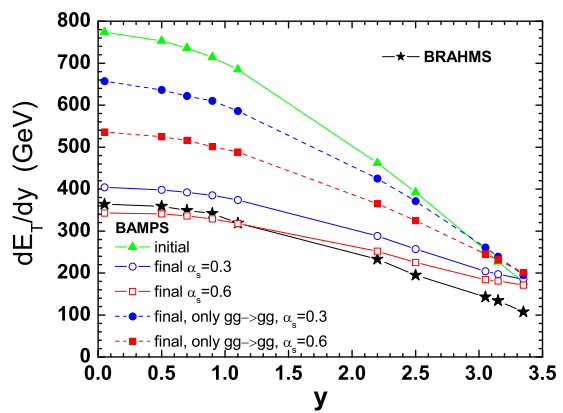}
\end{center}
\vspace{-0.6cm}
\caption{(color online). Rapidity dependence of the transverse energy
per rapidity. Results from BAMPS are multiplied by a factor of $2/3$
to compare with the BRAHMS data for charged mesons.
}
\label{fig:ety}
\end{figure}

Figure \ref{fig:ety} shows the rapidity dependence of the initial and final
$dE_T/dy$ (solid curves with open symbols) for the most central $5\%$
collisions, compared with
the BRAHMS data for the charged meson rapidity distributions,
$dE_T/dy=m_T\, dN/dy$ \cite{brahms}.
The shape of $dE_T/dy$ obtained from BAMPS agrees well with
the BRAHMS data within $y < 2.5$. 

The chosen initial condition of minijets, with $p_0=1.4$ GeV as a parameter,
seems appropriate for describing the experimental data.
However, calculations with the same initial condition and employing only
elastic pQCD $gg\to gg$ interactions show much less decrease of  $dE_T/dy$
\cite{XG05} (see dashed curves with solid symbols in Fig. \ref{fig:ety}).

\begin{figure}[ht]
\begin{center}
\includegraphics[height=5.6cm]{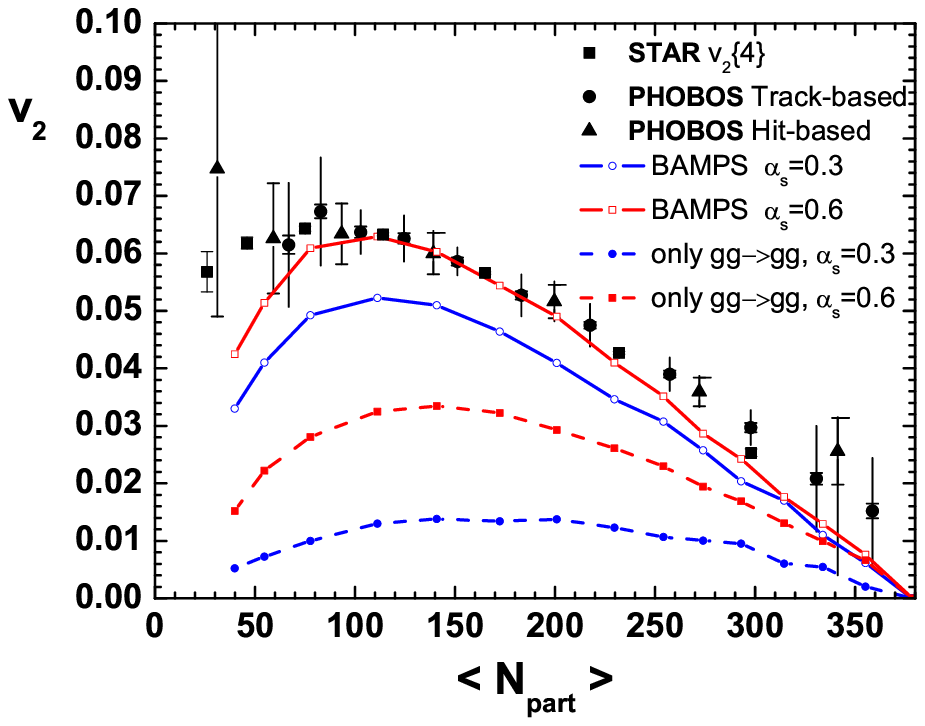}
\vfill
\includegraphics[height=9.8cm]{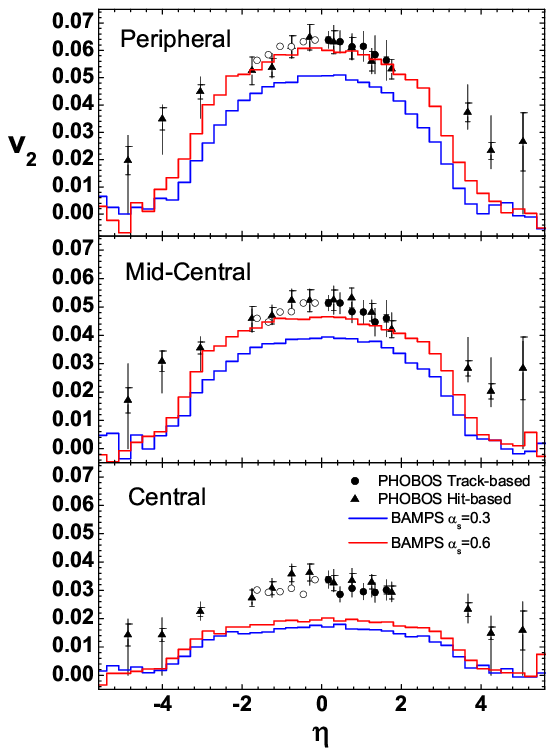}
\end{center}
\vspace{-0.6cm}
\caption{(color online). Upper panel: Elliptic flow $v_2(|y|<1)$ from BAMPS
using $\alpha_s=0.3$ and $0.6$, compared with the RHIC data \cite{phobos2}.
Lower panels: Pseudorapidity dependence of elliptic flow for three centrality
classes ranging from peripheral to central ($25\%-50\%$, $15\%-25\%$, and
$0\%-15\%$) from top to bottom.
}
\label{fig:v2}
\end{figure}

Figure \ref{fig:v2} shows the elliptic flow $v_2$ at midrapidity and
its dependence on rapidity, as obtained from the BAMPS calculations with
$\alpha_s=0.3,$ and $0.6$. These are
compared with the PHOBOS data \cite{phobos2}.

Except for the central centrality region the results with $\alpha_s=0.6$
agree perfectly with the experimental data, whereas the results with
$\alpha_s=0.3$ are roughly $20\%$ smaller. 
The results with only $gg\to gg$ interactions (dashed curves with solid
symbols) are much smaller than the data, which demonstrates that
bremsstrahlung and backreactions are the dominant processes in
generating large elliptic flow.

\begin{figure}[ht]
\begin{center}
\includegraphics[height=5.2cm]{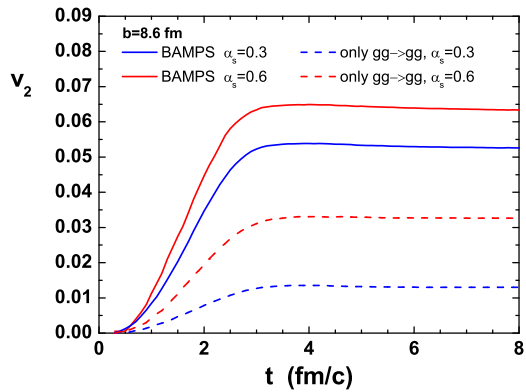}
\end{center}
\vspace{-0.6cm}
\caption{(color online). Generation of elliptic flow at midrapidity
for a noncentral collision with an impact parameter of $b=8.6$ fm.
}
\label{fig:v2t}
\end{figure}

Figure \ref{fig:v2t} shows the generation of the elliptic flow at midrapidity
for noncentral Au+Au collisions with impact parameter $b=8.6$ fm
corresponding to $\langle N_{\rm part} \rangle=111$.

$v_2$ increases strongly during the early stage of thermal equilibration 
(time scale $\stackrel{<}{\sim} 1$ fm/c) where hydrodynamics is
not valid but where the eccentricity is largest. The generation rate of
$v_2$ during thermalization is only slightly smaller than that during
the subsequent (viscous) hydrodynamical expansion. The $v_2$ generation
ends at $3$ fm/c, where the critical energy density is reached for all
the gluons at midrapidity.

Hence, the generation of the large elliptic flow observed at RHIC is well
described by pure perturbative gluon interactions as incorporated in BAMPS.
It is consistent with the decrease of the transverse energy due to the
mechanical work done by the pressure during the expansion. Inclusion of
hadronization \cite{MV03} and subsequent hadronic cascade \cite{ZBS05}
will yield only moderate contributions to the final
elliptic flow values, as the densities are already rather low and
the eccentricity becomes small.

The ratio of the shear viscosity to the entropy density, $\eta/s$, is
extracted from the BAMPS calculations (including $gg\leftrightarrow ggg$)
to see how viscous the gluon matter is for generating
the large $v_2$ values that match the experimental data.
We obtain the shear viscosity as \cite{XGviscos}
\begin{equation}
\label{viscos}
\eta \cong \frac{1}{5} n \frac{\langle E(\frac{1}{3}-v_z^2) \rangle}
{\frac{1}{3}-\langle v_z^2 \rangle} \frac{1}{\sum R^{\rm tr}+ 
\frac{3}{2} R_{23}-R_{32}}\,,
\end{equation}
where $n$ is the gluon density, $E$ is the gluon energy, and $v_z=p_z/E$
is the velocity in the beam direction. $\sum R^{\rm tr}$ denotes the
total transport collision rate \cite{XG07} describing momentum
isotropization. $R_{23}$ and $R_{32}$ are interaction rates for
$gg\to ggg$ and its backreaction, respectively. The entropy density
is extracted by assuming that the gluon matter is in local
kinetic equilibrium, which gives $s=4n-n\,\ln \lambda$, where 
$\lambda=n/n_{\rm eq}$ defines the gluon fugacity. The true entropy
density is expected to be (slightly) smaller, because overall kinetic
equilibration cannot be complete in an expanding system.

\begin{figure}[b]
\begin{center}
\includegraphics[height=5.2cm]{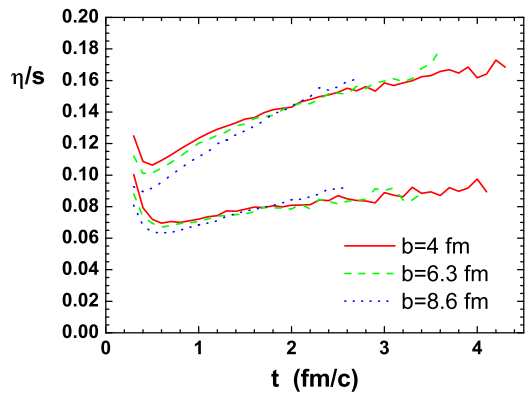}
\end{center}
\vspace{-0.6cm}
\caption{(color online). Shear viscosity to entropy density ratio
$\eta/s$ at the central region. The upper band shows the results with
$\alpha_s=0.3$ and the lower band the results with $\alpha_s=0.6$.
}
\label{fig:eos}
\end{figure}

Figure \ref{fig:eos} depicts the shear viscosity over entropy density ratio
$\eta/s$ extracted locally at the central region during the entire expansion
before decoupling.

The central region is the region with transverse radius of
$1.5$ fm and within the space-time rapidity interval $|\eta_t| < 0.2$,
where $\eta_t=\frac{1}{2} \ln [ (t+z)/(t-z) ]$. The curves in 
Fig.\ref{fig:eos} are obtained from BAMPS simulations with impact 
parameter $b=4$, $6.3$, and $8.6$ fm, for $\alpha_s=0.3$ and $0.6$.
Note that during thermal equilibration (time scale $\stackrel{<}{\sim} 1$ fm/c)
Eq. (\ref{viscos}) for calculating the shear viscosity is not fully
valid and the true entropy density will be smaller than that given by
the formula used.

The $\eta/s$ ratio obtained from the simulations with the same
$\alpha_s$ and for various impact parameters are almost the same,
although the gluon density
with $b=8.6$ fm is significantly smaller than that with $b=4$ fm.
Interaction rates and transport collision rates
scale with the temperature $T$. The gluon fugacity is close to $1$.
Hence, $\eta/s$ depends practically only on $\alpha_s$. This also explains
that the $\eta/s$ values extracted in the outer transverse regions (not shown)
are nearly the same as the ratio in the center.
Figure \ref{fig:eos} shows $\eta/s \approx 0.15$ for $\alpha_s=0.3$
and $\eta/s \approx 0.08$ for $\alpha_s=0.6$. These values agree well
with those found for a static thermal gluon gas \cite{XGviscos}.

\begin{figure}[h]
\begin{center}
\includegraphics[height=6cm]{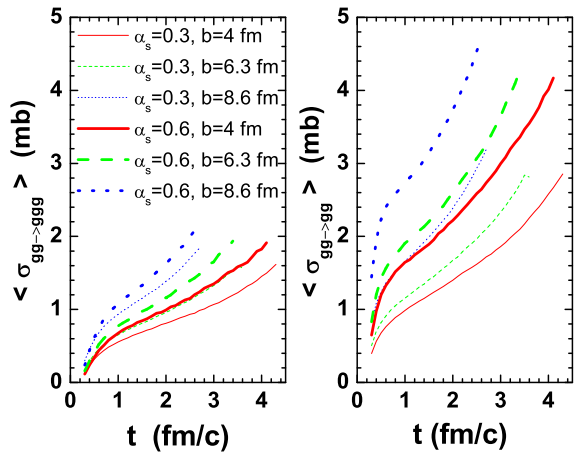}
\end{center}
\vspace{-0.6cm}
\caption{(color online). Cross sections of $gg\to gg$ and $gg \to ggg$
processes in the central region.
}
\label{fig:cs}
\end{figure}

Figure \ref{fig:cs} shows the cross sections of $gg\to gg$ and
$gg \to ggg$ processes, extracted in the central region.

Indeed, the cross section becomes larger when the density is smaller, so that
$\langle \sigma \rangle \, T^2$ depends practically only on $\alpha_s$.
Note that
the cross sections are indeed small. This demonstrates that
it is not necessary to use unphysically large cross sections \cite{MG02}
or ultrashort hadronization times \cite{BS02} to yield
large $v_2$ values.

The pQCD based parton cascade BAMPS is used to calculate the
elliptic flow $v_2$ and to extract the ratio of the shear viscosity to
the entropy density, $\eta/s$, from simulations of Au+Au collisions at
RHIC energy $\sqrt{s}=200$A GeV. Simple parton-hadron duality allows one
to compare our final $v_2$ results with the data measured at RHIC.
Agreement between the data and theory is found with Glauber-type minijets
initial conditions for gluons and with $\alpha_s=0.3-0.6$. The $\eta/s$ 
ratio of the gluon plasma created varies between $0.15$ and $0.08$.
Standard pQCD interactions alone can describe the generation of large
$v_2$ values at RHIC. The small $\eta/s$ ratios found in
the simulations indicate that the gluon plasma created behaves
like a nearly perfect fluid. This can be understood by perturbative QCD
without resorting to exotic explanations such as
the AdS/CFT conjecture: Gluon bremsstrahlung dominates and
yields rapid thermalization, and therefore early pressure buildup
and a small shear viscosity in the gluon gas.

\acknowledgments

The authors thank G.~Burau and M.~Gyulassy for fruitful
discussions.

\end{document}